\documentclass[reprint, superscriptaddress, amsmath,amssymb, aps, prd]{revtex4-2}

\usepackage{graphics}
\usepackage{graphicx}
\usepackage{dcolumn}
\usepackage{bm}
\usepackage{amsmath}
\usepackage{amssymb}
\usepackage{xcolor}
\usepackage{dsfont}
\usepackage{braket}
\usepackage{stfloats}
\usepackage{hyperref}
\numberwithin{equation}{section}
\usepackage{lineno}

\begin{document}


\title{Searching for Extra Dimensions and Copies of the Standard Model with IceCube}


\affiliation{III. Physikalisches Institut, RWTH Aachen University, D-52056 Aachen, Germany}
\affiliation{Department of Physics, University of Adelaide, Adelaide, 5005, Australia}
\affiliation{Dept. of Physics and Astronomy, University of Alaska Anchorage, 3211 Providence Dr., Anchorage, AK 99508, USA}
\affiliation{School of Physics and Center for Relativistic Astrophysics, Georgia Institute of Technology, Atlanta, GA 30332, USA}
\affiliation{Dept. of Physics, Southern University, Baton Rouge, LA 70813, USA}
\affiliation{Dept. of Physics, University of California, Berkeley, CA 94720, USA}
\affiliation{Lawrence Berkeley National Laboratory, Berkeley, CA 94720, USA}
\affiliation{Institut f{\"u}r Physik, Humboldt-Universit{\"a}t zu Berlin, D-12489 Berlin, Germany}
\affiliation{Fakult{\"a}t f{\"u}r Physik {\&} Astronomie, Ruhr-Universit{\"a}t Bochum, D-44780 Bochum, Germany}
\affiliation{Universit{\'e} Libre de Bruxelles, Science Faculty CP230, B-1050 Brussels, Belgium}
\affiliation{Vrije Universiteit Brussel (VUB), Dienst ELEM, B-1050 Brussels, Belgium}
\affiliation{Dept. of Physics, Simon Fraser University, Burnaby, BC V5A 1S6, Canada}
\affiliation{Department of Physics and Laboratory for Particle Physics and Cosmology, Harvard University, Cambridge, MA 02138, USA}
\affiliation{Dept. of Physics, Massachusetts Institute of Technology, Cambridge, MA 02139, USA}
\affiliation{Dept. of Physics and The International Center for Hadron Astrophysics, Chiba University, Chiba 263-8522, Japan}
\affiliation{Department of Physics, Loyola University Chicago, Chicago, IL 60660, USA}
\affiliation{Dept. of Physics and Astronomy, University of Canterbury, Private Bag 4800, Christchurch, New Zealand}
\affiliation{Dept. of Physics, University of Maryland, College Park, MD 20742, USA}
\affiliation{Dept. of Astronomy, Ohio State University, Columbus, OH 43210, USA}
\affiliation{Dept. of Physics and Center for Cosmology and Astro-Particle Physics, Ohio State University, Columbus, OH 43210, USA}
\affiliation{Niels Bohr Institute, University of Copenhagen, DK-2100 Copenhagen, Denmark}
\affiliation{Dept. of Physics, TU Dortmund University, D-44221 Dortmund, Germany}
\affiliation{Dept. of Physics and Astronomy, Michigan State University, East Lansing, MI 48824, USA}
\affiliation{Dept. of Physics, University of Alberta, Edmonton, Alberta, T6G 2E1, Canada}
\affiliation{Erlangen Centre for Astroparticle Physics, Friedrich-Alexander-Universit{\"a}t Erlangen-N{\"u}rnberg, D-91058 Erlangen, Germany}
\affiliation{Physik-department, Technische Universit{\"a}t M{\"u}nchen, D-85748 Garching, Germany}
\affiliation{D{\'e}partement de physique nucl{\'e}aire et corpusculaire, Universit{\'e} de Gen{\`e}ve, CH-1211 Gen{\`e}ve, Switzerland}
\affiliation{Dept. of Physics and Astronomy, University of Gent, B-9000 Gent, Belgium}
\affiliation{Dept. of Physics and Astronomy, University of California, Irvine, CA 92697, USA}
\affiliation{Karlsruhe Institute of Technology, Institute for Astroparticle Physics, D-76021 Karlsruhe, Germany}
\affiliation{Karlsruhe Institute of Technology, Institute of Experimental Particle Physics, D-76021 Karlsruhe, Germany}
\affiliation{Dept. of Physics, Engineering Physics, and Astronomy, Queen's University, Kingston, ON K7L 3N6, Canada}
\affiliation{Department of Physics {\&} Astronomy, University of Nevada, Las Vegas, NV 89154, USA}
\affiliation{Nevada Center for Astrophysics, University of Nevada, Las Vegas, NV 89154, USA}
\affiliation{Dept. of Physics and Astronomy, University of Kansas, Lawrence, KS 66045, USA}
\affiliation{UCLouvain, Centre for Cosmology, Particle Physics and Phenomenology, CP3, Chemin du Cyclotron 2, 1348 Louvain-la-Neuve, Belgium}
\affiliation{Department of Physics, Mercer University, Macon, GA 31207-0001, USA}
\affiliation{Dept. of Astronomy, University of Wisconsin{\textemdash}Madison, Madison, WI 53706, USA}
\affiliation{Dept. of Physics and Wisconsin IceCube Particle Astrophysics Center, University of Wisconsin{\textemdash}Madison, Madison, WI 53706, USA}
\affiliation{Institute of Physics, University of Mainz, Staudinger Weg 7, D-55099 Mainz, Germany}
\affiliation{Department of Physics, Marquette University, Milwaukee, WI 53201, USA}
\affiliation{Institut f{\"u}r Kernphysik, Universit{\"a}t M{\"u}nster, D-48149 M{\"u}nster, Germany}
\affiliation{Bartol Research Institute and Dept. of Physics and Astronomy, University of Delaware, Newark, DE 19716, USA}
\affiliation{Dept. of Physics, Yale University, New Haven, CT 06520, USA}
\affiliation{Columbia Astrophysics and Nevis Laboratories, Columbia University, New York, NY 10027, USA}
\affiliation{Dept. of Physics, University of Oxford, Parks Road, Oxford OX1 3PU, United Kingdom}
\affiliation{Dipartimento di Fisica e Astronomia Galileo Galilei, Universit{\`a} Degli Studi di Padova, I-35122 Padova PD, Italy}
\affiliation{Dept. of Physics, Drexel University, 3141 Chestnut Street, Philadelphia, PA 19104, USA}
\affiliation{Physics Department, South Dakota School of Mines and Technology, Rapid City, SD 57701, USA}
\affiliation{Dept. of Physics, University of Wisconsin, River Falls, WI 54022, USA}
\affiliation{Dept. of Physics and Astronomy, University of Rochester, Rochester, NY 14627, USA}
\affiliation{Department of Physics and Astronomy, University of Utah, Salt Lake City, UT 84112, USA}
\affiliation{Dept. of Physics, Chung-Ang University, Seoul 06974, Republic of Korea}
\affiliation{Oskar Klein Centre and Dept. of Physics, Stockholm University, SE-10691 Stockholm, Sweden}
\affiliation{Dept. of Physics and Astronomy, Stony Brook University, Stony Brook, NY 11794-3800, USA}
\affiliation{Dept. of Physics, Sungkyunkwan University, Suwon 16419, Republic of Korea}
\affiliation{Institute of Physics, Academia Sinica, Taipei, 11529, Taiwan}
\affiliation{Dept. of Physics and Astronomy, University of Alabama, Tuscaloosa, AL 35487, USA}
\affiliation{Dept. of Astronomy and Astrophysics, Pennsylvania State University, University Park, PA 16802, USA}
\affiliation{Dept. of Physics, Pennsylvania State University, University Park, PA 16802, USA}
\affiliation{Dept. of Physics and Astronomy, Uppsala University, Box 516, SE-75120 Uppsala, Sweden}
\affiliation{Dept. of Physics, University of Wuppertal, D-42119 Wuppertal, Germany}
\affiliation{Deutsches Elektronen-Synchrotron DESY, Platanenallee 6, D-15738 Zeuthen, Germany}

\author{R. Abbasi}
\affiliation{Department of Physics, Loyola University Chicago, Chicago, IL 60660, USA}
\author{M. Ackermann}
\affiliation{Deutsches Elektronen-Synchrotron DESY, Platanenallee 6, D-15738 Zeuthen, Germany}
\author{J. Adams}
\affiliation{Dept. of Physics and Astronomy, University of Canterbury, Private Bag 4800, Christchurch, New Zealand}
\author{J. A. Aguilar}
\affiliation{Universit{\'e} Libre de Bruxelles, Science Faculty CP230, B-1050 Brussels, Belgium}
\author{M. Ahlers}
\affiliation{Niels Bohr Institute, University of Copenhagen, DK-2100 Copenhagen, Denmark}
\author{J.M. Alameddine}
\affiliation{Dept. of Physics, TU Dortmund University, D-44221 Dortmund, Germany}
\author{S. Ali}
\affiliation{Dept. of Physics and Astronomy, University of Kansas, Lawrence, KS 66045, USA}
\author{N. M. Amin}
\affiliation{Bartol Research Institute and Dept. of Physics and Astronomy, University of Delaware, Newark, DE 19716, USA}
\author{K. Andeen}
\affiliation{Department of Physics, Marquette University, Milwaukee, WI 53201, USA}
\author{C. Arg{\"u}elles}
\affiliation{Department of Physics and Laboratory for Particle Physics and Cosmology, Harvard University, Cambridge, MA 02138, USA}
% ...
\date{\today}

\collaboration{IceCube Collaboration}
\noaffiliation

\begin{abstract}
The hierarchy problem remains an open question in particle physics. A number of theories that address this problem lower the fundamental scale of gravity, resulting in observable consequences in the neutrino sector. In this work, we place constraints on low-scale gravity scenarios using high-energy neutrinos observed with the IceCube Neutrino Observatory. The analysis is based on 10.7 years of upward-going muon neutrino data in the energy range from 0.5 to 100 TeV.
In this energy range, the theories predict characteristic spectral distortions arising from matter effects when neutrinos propagate through Earth.
In the context of large extra dimension models, we constrain the compactification radius of the largest extra dimension to $R \lesssim 0.17\,\mu\mathrm{m}$ at $90\%$ confidence level for both normal and inverted neutrino mass ordering. For scenarios with multiple Standard Model copies, we obtain lower limits of up to $N \gtrsim \mathcal{O}(400)$, depending on the value of the lightest neutrino mass. In parts of the parameter space, these results constitute the strongest constraints in the literature to our knowledge, while in other regions they probe previously unexplored parameter space.
\end{abstract}

\maketitle

\section{Introduction \label{sec: Introduction}} 

The hierarchy problem is one of the central open questions in modern particle physics. In the Standard Model (SM), quantum corrections to the Higgs boson mass scale quadratically with the ultraviolet cutoff of the theory. However, the Higgs boson was observed in 2012 at the Large Hadron Collider \cite{aad2012observation,CMS:2012qbp} with a mass of approximately $125\,\mathrm{GeV}$~\cite{ParticleDataGroup:2026aaa}. If the cutoff is taken to be the Planck scale, $M_P = 1.22 \times 10^{19}\,\mathrm{GeV}$, stabilizing the Higgs mass at the electroweak scale requires an extreme fine-tuning between the bare mass and radiative corrections. The enormous disparity between the electroweak scale and the gravitational scale therefore calls for an explanation.

Low-scale gravity theories offer an alternative approach to this problem by postulating that the fundamental scale of gravity, hereafter denoted as $M_f$, is not given by $M_P$, but is instead much lower, potentially as small as the TeV scale. By reducing the effective ultraviolet cutoff relevant for quantum corrections, these frameworks narrow the gap between the Higgs mass and the fundamental scale of the theory, thereby alleviating the hierarchy problem. In this work, we consider two benchmark realizations of this idea: the Arkani--Hamed--Dimopoulos--Dvali (ADD) model with $d$ compact extra spatial dimensions~\cite{arkani1998hierarchy,arkani1999phenomenology}, and the framework with $N$ copies of the SM~\cite{dvali2008black,dvali2009phenomenology}. While conceptually distinct, both scenarios lower the fundamental scale of gravity through the presence of additional degrees of freedom. In fact, both concepts are deeply connected \cite{Dvali:2010vm}.

Beyond addressing the hierarchy problem, these frameworks naturally connect to other open questions of the SM. In particular, they provide mechanisms to generate small neutrino masses \cite{Arkani-Hamed:1998wuz,Ettengruber:2022pxf} and can accommodate viable dark matter candidates \cite{Arkani-Hamed:1999rvc, Dvali:2009ne, Dvali:2009fw, Dvali:2019ewm}. In the neutrino sector, we extend both scenarios by introducing a Dirac right-handed neutrino that couples to the SM via Yukawa interactions. The presence of additional sterile states modifies neutrino propagation in matter. As neutrinos traverse the Earth, matter effects can enhance active--sterile transitions, leading to distortions in the energy and zenith-angle distributions measured at high-energy neutrino telescopes such as IceCube.

IceCube \cite{IceCube:2016zyt} is a cubic-kilometer-scale neutrino telescope located at the geographic South Pole. The detector instruments a volume of approximately $1\,\mathrm{km}^3$ of Antarctic ice with 86 vertical strings, each equipped with digital optical modules (DOMs), for a total of 5160 optical sensors \cite{Abbasi_2010}. These DOMs detect Cherenkov photons emitted by charged secondary particles produced in neutrino interactions occurring within or near the instrumented volume. This capability makes IceCube particularly sensitive to energy- and zenith-dependent distortions induced by matter-enhanced active--sterile transitions.

The absence of such distortions in data can therefore be translated into constraints on the physical parameters of low-scale gravity scenarios. In this paper, we exploit 10.7 years of high-energy neutrino data collected with the IceCube Neutrino Observatory to probe matter-enhanced transitions induced by these models and derive limits on the compactification radius in ADD and on the number of SM copies in the many-species framework.

\section{Models and their neutrino phenomenology}
\label{sec:model}

Low-scale gravity scenarios have been probed with sub-millimeter tests of the
gravitational force, with astrophysical and cosmological observables, and through
the absence of signals at colliders~\cite{ParticleDataGroup:2026aaa}. Here we
confront two of them with high-energy neutrinos.

In both models the neutrino sector is extended by a Dirac right-handed neutrino
per flavor, which is a gauge singlet and therefore couples to our models under test. This results in sterile neutrino oscillations, whose masses and mixing parameters are controlled by
the same parameters that lower the fundamental scale of gravity. Their phenomenology is closely related to that of the eV-scale sterile neutrino searches
performed with the same IceCube data sample~\cite{IceCubeCollaboration:2024nle,IceCubeCollaboration:2024dxk}: sterile states in the eV mass range induce matter-enhanced conversion of atmospheric
$\bar\nu_\mu$ as they traverse the Earth, which appears as a deficit localized in
energy and zenith angle. What differs is the structure of the sterile sector. A
$3+1$ search introduces a single sterile state whose mass splitting and mixing
angles are independent phenomenological parameters, motivated largely by
short-baseline anomalies \cite{Diaz:2019fwt, Boser:2019rta}. The models considered here instead predict several
sterile states at once, but their masses and mixing angles are not free parameters, but are dictated by the theory. The result is a scenario with more sterile states
but fewer free parameters, and with a motivation rooted in the hierarchy problem
rather than in an anomaly.

\subsection{Matter-enhanced conversion in the Earth}
\label{subsec:resonances}

\begin{figure}[h]
    \centering
    \includegraphics[width=\linewidth]{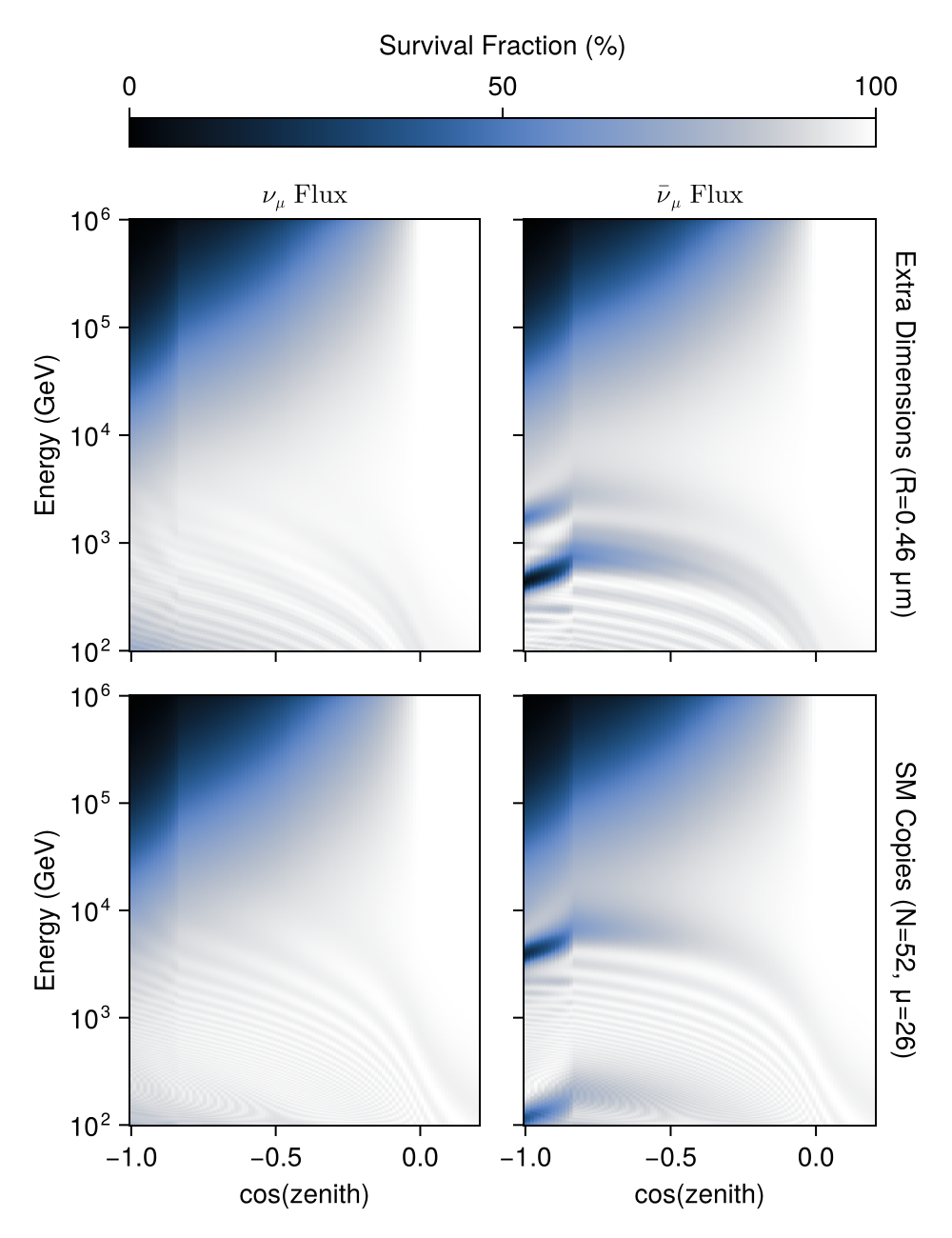}
    \caption{Matter-enhanced oscillation signatures in ADD (top) and SM copies (bottom).
    Relative flux distortion after propagation through the Earth as a function of true neutrino energy and zenith angle for representative parameter choices in the two models are shown as survival fractions of the initial flux. The resonant features present for anti-neutrinos illustrate the characteristic signal structure probed in this analysis. The top left corner of each plot shows significant Earth absorption of neutrinos. For the ADD scenario, we consider a compactification radius of $R = 0.46\,\mu\mathrm{m}$. For the SM copies framework, we adopt $N = 52$ and $\mu = 26$. In both cases, the lightest neutrino mass is set to $m_0 = 0\,\mathrm{eV}$.}
    \label{fig: comparison flux ratio models}
\end{figure}

The imprint of these models is the cleanest at the level of the incident flux, before detector effects are folded in. Figure~\ref{fig: comparison flux ratio models} shows the predicted flux, normalized to the SM expectation, in the plane of neutrino energy and the cosine of the zenith angle, for representative benchmark points of the ADD and SM-copies scenarios.

The zenith angle fixes the propagation baseline. Down-going neutrinos (cos(zenith) $ = 1$) reach the detector after crossing only a thin column of atmosphere and ice, whereas up-going neutrinos (cos(zenith) $ = -1$) traverse the full diameter of the Earth. The baseline thus grows monotonically as the trajectory tips from horizontal to vertical, and with it the column of matter the neutrinos cross. This matter is not uniform: the density follows the layered Preliminary Reference Earth Model (PREM) profile \cite{Dziewonski:1981xy}, rising gradually through the mantle and then jumping sharply at the core–mantle boundary. Trajectories begin to graze the outer core near
cos(zenith) $ \approx -0.84$, and this density discontinuity is inherited by the oscillation pattern, appearing as the band-splitting that separates mantle-only from core-crossing paths in Fig.~\ref{fig: comparison flux ratio models}.

The large flux deficit appearing at the upper left corner in all panels of Fig.~\ref{fig: comparison flux ratio models} is due to Earth absorption \cite{IceCube:2017roe}. The more localized distortions in the anti-neutrinos---our signal---originate from matter-enhanced flavor conversions. 
In the familiar solar and atmospheric matter effects, only the charged-current (CC) potential of $\nu_e$ is physical, while the neutral-current (NC) one is flavor-symmetric and reduces to an unobservable overall phase. However, here it is the NC potential that matters: all active flavors acquire it from coherent scattering off nucleons, but sterile neutrinos do not, so it survives as a physical active–sterile potential difference. Because this potential is negative, and because the sterile states in both models are heavier than the active ones, the Mikheyev–Smirnov–Wolfenstein (MSW) \cite{Wolfenstein:1977ue, Mikheyev:1985zog} resonance condition can only be satisfied for antineutrinos.
This is the origin of the $\bar\nu$-only structure seen in Fig.~\ref{fig: comparison flux ratio models}. The condition relates the resonance energy to the active-sterile mass splitting,
\begin{equation*}
  E_{\rm res} \;\sim\; \mathcal{O}(1)\;\mathrm{TeV}
  \;\times\; \frac{\Delta m^2}{1\,\mathrm{eV}^2},
  \label{eq:Eres}
\end{equation*}
with an $\mathcal{O}(1)$ coefficient that depends on the matter density along the trajectory, and hence on the zenith angle. Denser, core-crossing paths push the resonance to lower energies, producing the bending of the resonant bands with zenith angle in Fig.~\ref{fig: comparison flux ratio models}, while the alternating mantle–core–mantle layers along these paths give an undulating matter potential that parametrically enhances the conversion~\cite{Akhmedov:1999ty}.
The energy range of our sample, 0.5 to 100~TeV, therefore selects sterile masses at the eV scale, the same range probed by the $3+1$ analysis.

Two further features carry over from that analysis. The depth of the dip is governed by the active-sterile mixing angle, so that the sensitivity of the present search is inherited directly from the sensitivity of the sterile search to $\sin^2 2\theta$. And because the conversion builds up along the trajectory, the distortion is strongest for the most vertical events. Each model therefore has to answer only two questions: where do the resonances sit in energy, and how deep are they?

\subsection{Extra dimensions (ADD)}
\label{subsec:add}

The ADD framework~\cite{arkani1998hierarchy,arkani1999phenomenology} postulates
$d$ compact extra dimensions, whose volume suppresses the fundamental scale of gravity
relative to the Planck scale ($M_p$). The right-handed neutrino can propagate in the bulk and
acquires a Dirac mass suppressed by the same volume factor, which explains the
smallness of the active neutrino masses~\cite{Dvali:1999cn,Arkani-Hamed:1998wuz}.
From the brane point of view, it appears as an infinite tower of Kaluza-Klein (KK)
modes, so that every light mass eigenstate is accompanied by a series of sterile
partners. The construction and diagonalization of the mass matrix are given in
Refs.~\cite{Dvali:1999cn,Ettengruber:2022pxf}; here we only need its two
ingredients, the KK masses and their mixing with the active states,
\begin{equation*}
  m^{(n)} \simeq \frac{n}{R},
  \qquad
  \theta_i^{(n)} \simeq \frac{\sqrt{2}\, m_i R}{n},
  \qquad n = 1,2,\dots
  \label{eq:kk}
\end{equation*}
where $R$ is the radius of the largest extra dimension and $m_i$ the light
neutrino masses.

The model therefore has only two parameters, $R$ and the lightest neutrino mass
$m_0$, and each plays a distinct role. The radius sets the sterile mass scale.
Since the KK masses grow linearly with $n$, the mass splittings grow as $n^2/R^2$
and the tower produces a whole ladder of resonances rather than a single one,
spaced quadratically in energy. Decreasing $R$ makes the sterile states heavier and
pushes the entire ladder up in energy. Sub-micrometer radii place the lowest KK
resonances in the TeV range, which is precisely the region covered by our sample;
for radii much above a micrometer the resonances drop below the energy threshold, but those are already excluded by direct tests of the gravitational force~\cite{ParticleDataGroup:2026aaa}. For very small radii, the resonances move beyond the reach of the available statistics, meaning sub-micrometer radii are the relevant range, the benchmark $R = 0.46\,\mu\mathrm{m}$ used in Fig.~\ref{fig: comparison flux ratio models}
lies in this range and is chosen so that the first two resonances fall inside the analysis window, making the ladder structure clearly visible.
The lightest neutrino mass, on the other hand, does not significantly move the resonances, but it affects the mixing, which is proportional to the light neutrino masses and $R$. A heavier active spectrum, therefore, leaves the ladder of resonances where it is and simply makes each dip deeper, while smaller radii also suppress the mixing.

The higher KK modes matter progressively less: their resonances move to
higher energies as $n^2$, where the atmospheric flux is steeply falling, while
their mixing decreases as $1/n$. In practice, only the lowest modes are relevant,
which is what allows the tower to be truncated (Sec.~\ref{sec:analysis}).

\subsection{Standard Model copies}
\label{subsec:copies}

In the many-species framework, the SM is replicated $N$
times~\cite{dvali2008black,dvali2009phenomenology}, with the copies interacting with one
another only gravitationally, and the large number of species lowers the
fundamental scale through the species bound. Since the right-handed neutrino is a
singlet, it couples to all copies, and permutation symmetry restricts the Yukawa
structure to two independent
couplings~\cite{dvali2009phenomenology,ettengruber2024testing}. Diagonalizing the
resulting mass matrix gives, for each flavor, $N-1$ degenerate light states and a
single heavy state, the symmetric combination of all copies. There is no tower
here like in the ADD case, and our neutrino communicates with the other copies through the single heavy state,
with which it has an overlap $1/\sqrt{N}$. Each flavor therefore behaves as an effective
two-level system, with
\begin{equation*}
  \Delta m_i^2 = (\mu^2 - 1)\, m_i^2 ,
  \qquad
  \sin^2 2\theta \simeq \frac{4}{N},
  \label{eq:copies}
\end{equation*}
where the mass factor $\mu$ is the ratio of the heavy to the light mass, assumed to be
flavor universal.

This means that three parameters, the number of copies $N$, the mass factor $\mu$, and the lightest neutrino mass $m_0$  determine the observable distortions in the neutrino spectrum. The
mass factor $\mu$ sets where the resonances sit: it lifts the heavy states above
the light ones, and only when it brings them to the eV scale do the resonances fall
into the energy range of the sample. Since the light masses are of order tens of
meV, this requires $\mu$ of order tens, and the accessible window in $\mu$ is
correspondingly narrow. The number of copies $N$ controls the mixing strength and therefore sets how deep the dips are. The reach in $N$ is thus set directly by the sensitivity of the sterile search to $\sin^2 2\theta$, and a sensitivity at the percent level corresponds to $N$ of
order a few hundred. Finally, the lightest neutrino mass $m_0$ determines how the
three resonances are distributed.
The benchmark $N = 52$, $\mu = 26$ in Fig.~\ref{fig: comparison flux ratio models} is chosen accordingly: it brings the heavy states to the eV scale, placing the $m_3$ resonance at a few TeV where the sample is most sensitive, while the mixing $\sin^2 2\theta \simeq 4/N \approx 0.08$ is large enough for the distortion to be clearly visible.


\section{Analysis}
\label{sec:analysis}

\begin{figure}[h]
    \centering
    \includegraphics[width=\linewidth]{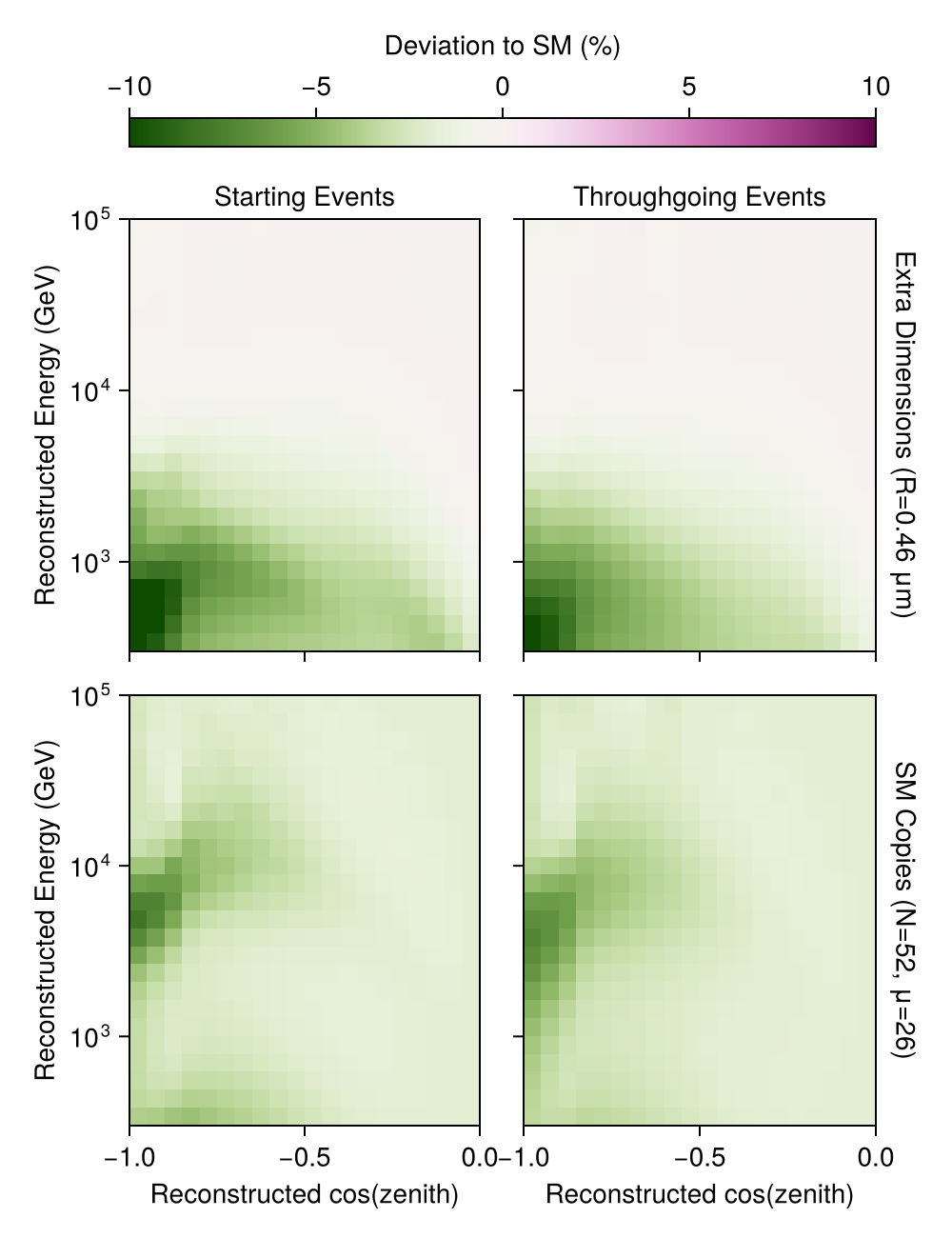}
    \caption{Expected signal signatures in reconstructed space.
Relative deviations (in percent) with respect to the SM expectation in reconstructed energy and zenith angle for representative benchmark parameters (same as Fig~\ref{fig: comparison flux ratio models}) of the ADD (top) and SM copies (bottom) scenarios. The left panels correspond to starting events and the right panels to through-going events.}
    \label{fig:reco_distortions}
\end{figure}

The analysis is based on data collected with the IceCube Neutrino Observatory, using the same event sample that had previously been developed and used for the 3+1 eV-scale sterile search \cite{IceCubeCollaboration:2024nle, IceCubeCollaboration:2024dxk}.
We focus on upward-going, track-like events originating from CC interactions of muon neutrinos, which provide good angular resolution and sensitivity to matter effects along long propagation baselines through the Earth. IceCube does not have the ability to separate $\nu_\mu$ from $\bar{\nu}_\mu$, and our data sample is a mixture of both types, while the matter resonances only appear in $\bar{\nu}_\mu$, and $\nu_\mu$ constitute a background. 
The data sample comprises 368{,}071 events recorded between May 2011 and June 2022, corresponding to 10.7~years of effective detector livetime, with an estimated purity of 99.91\% muon neutrino CC interactions. Due to differences in flux and interaction cross sections, the sample is dominated by roughly 74\% $\nu_\mu$, and $\bar{\nu}_\mu$ make up only around 26\% of the events.
The selected sample covers reconstructed neutrino energies from 0.5 to 100 TeV, where matter-enhanced active--sterile transitions may occur in the scenarios considered. Moreover, we consider two event topologies: starting events, where the neutrino interaction vertex is contained within the detector volume, and through-going events, where muons produced outside the detector traverse the instrumented region. In our dataset, 93{,}762 events are classified as starting and 274{,}309 as through-going tracks.
Figure~\ref{fig:reco_distortions} illustrates the expected relative deviations with respect to the SM in reconstructed space for representative benchmark parameters of the ADD and SM copies scenarios. As shown, starting tracks have better energy resolution, resulting in more defined resonant features. At the same time, the through-going tracks class benefits from higher event counts and thus smaller statistical uncertainties.

The expected event distributions are obtained through a forward-folding procedure. Atmospheric neutrino fluxes are modeled using calculations that include conventional, prompt, and astrophysical components. Neutrino propagation through the Earth is computed by numerically solving the evolution equation in matter with the \texttt{nuSQuIDS} framework~\cite{arguelles2022nusquids}, incorporating the PREM density profile. In the presence of additional sterile states, the Hamiltonian is extended accordingly, and oscillation probabilities are evaluated as a function of true neutrino energy and zenith angle.

For the ADD scenario the KK tower has to be truncated. As discussed in
Sec.~\ref{subsec:add}, the higher modes resonate at energies growing as $n^2$,
where the atmospheric flux is steeply falling, while their mixing to the active
states decreases as $1/n$. We retain the first two, dominant modes, which together with the
three light states result in a ($9 \times 9$) Hamiltonian, the largest dimension
supported by \texttt{nuSQuIDS}. Each additional mode adds a further resonance and
can therefore only increase the predicted disappearance, so that the truncation is
conservative. For the SM copies scenario, there is no truncation necessary, since the full mixing is described by a ($6 \times 6$) matrix.

The oscillated fluxes are then convolved with detector response functions derived from detailed simulations to obtain predictions in reconstructed energy and zenith angle. The full statistical analysis is performed within the \texttt{GollumFit} framework~\cite{icecube2025gollumfit}.

Detector effects are included using detailed Monte Carlo simulations of neutrino interactions, secondary particle propagation, Cherenkov light emission, and photon detection in the instrumented ice. \cite{IceCubeCollaboration:2024dxk} The simulated events are processed with the same reconstruction and selection algorithms as applied to data. The final analysis is performed in reconstructed energy and zenith-angle bins, which capture the characteristic distortions induced by matter-enhanced transitions.

Constraints on the model parameters are derived using a binned Poisson likelihood that is adjusted to account for the finite statistical uncertainties in Monte Carlo events used to construct the expected distributions \cite{arguelles2019binned}. 

Sources of systematic uncertainty are taken into account, following the same prescription as detailed in \cite{IceCubeCollaboration:2024nle, IceCubeCollaboration:2024dxk}. The treatment includes nuisance parameters with appropriate priors that affect the overall event normalization, the primary cosmic-ray spectrum, hadronic interaction modeling, non-conventional flux and its spectrum, interaction cross-sections, and detector uncertainties, including the optical properties of the South Pole glacier.
The likelihood is profiled over these nuisance parameters, and exclusion limits are constructed using the profile likelihood ratio and Wilks' theorem.

A more detailed description of the event sample used, including the distributions of the observed data, details on the event selection and reconstruction methods, and additional information on the nuisance parameters, can be found in \cite{IceCubeCollaboration:2024dxk}.

\section{Results}
\label{sec:results}

We find no significant deviation from the three-flavor neutrino oscillation framework. The best-fit points in both the ADD and SM copies parameter spaces are consistent with the null hypothesis within statistical fluctuations (significance $\ll3\sigma$). Therefore, we proceed to draw exclusion limits on the physical model parameters of our theories.

\subsection{Extra Dimensions (ADD)}

\begin{figure}[h]
    \centering
    \includegraphics[width=0.8\linewidth]{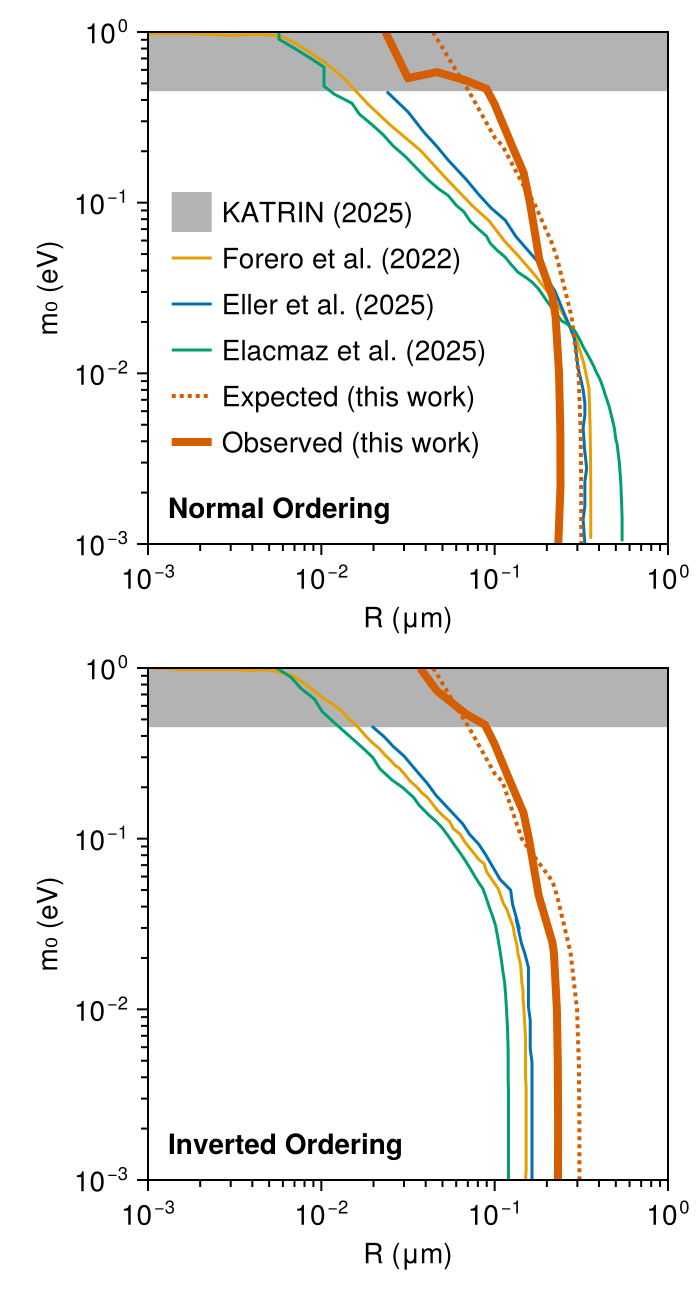}
    \caption{Upper limits for the radius of the largest extra dimension in ADD. Regions to the right of the lines are excluded at the 99\% confidence levels, shown separately for normal (top) and inverted (bottom) mass ordering. The solid red lines indicate our observed limits, while the dotted red lines indicate our median expected sensitivity. The remaining lines compare with existing neutrino bounds~\cite{eller2025neutrino,elaccmaz2025updated,forero2022large,katrin2025direct}.}
    \label{fig:add_limits}
\end{figure}

Figure~\ref{fig:add_limits} shows the exclusion limits derived for the ADD scenario in the plane of the compactification radius $R$ and the lightest neutrino mass $m_0$, for both normal and inverted mass ordering. At $90\%$ confidence level, we constrain the compactification radius of the largest extra dimension down to
\begin{equation*}
R \lesssim 0.17\,\mu\mathrm{m},
\end{equation*}
independently of the lightest neutrino mass $m_0$ and the neutrino mass ordering. The sensitivity weakly depends on the assumed mass ordering, with comparable limits obtained in both cases. For increasing values of $m_0$, the effective active--sterile mixing is enhanced, leading to stronger distortions in the event distributions and consequently improved constraints on the model parameters. Also shown in Fig.~\ref{fig:add_limits} are bounds from other neutrino measurements.
As shown there, our limit is competitive with or stronger than existing neutrino-oscillation constraints~\cite{eller2025neutrino,elaccmaz2025updated,forero2022large} over large parts of the $m_0$ range, and is complementary to the direct neutrino-mass bound from the Karlsruhe Tritium Neutrino Experiment (KATRIN)~\cite{katrin2025direct}, which constrains $m_0$ rather than $R$.

\subsection{Standard Model Copies}

\begin{figure*}[h]
    \centering
    \includegraphics[width=0.8\textwidth]{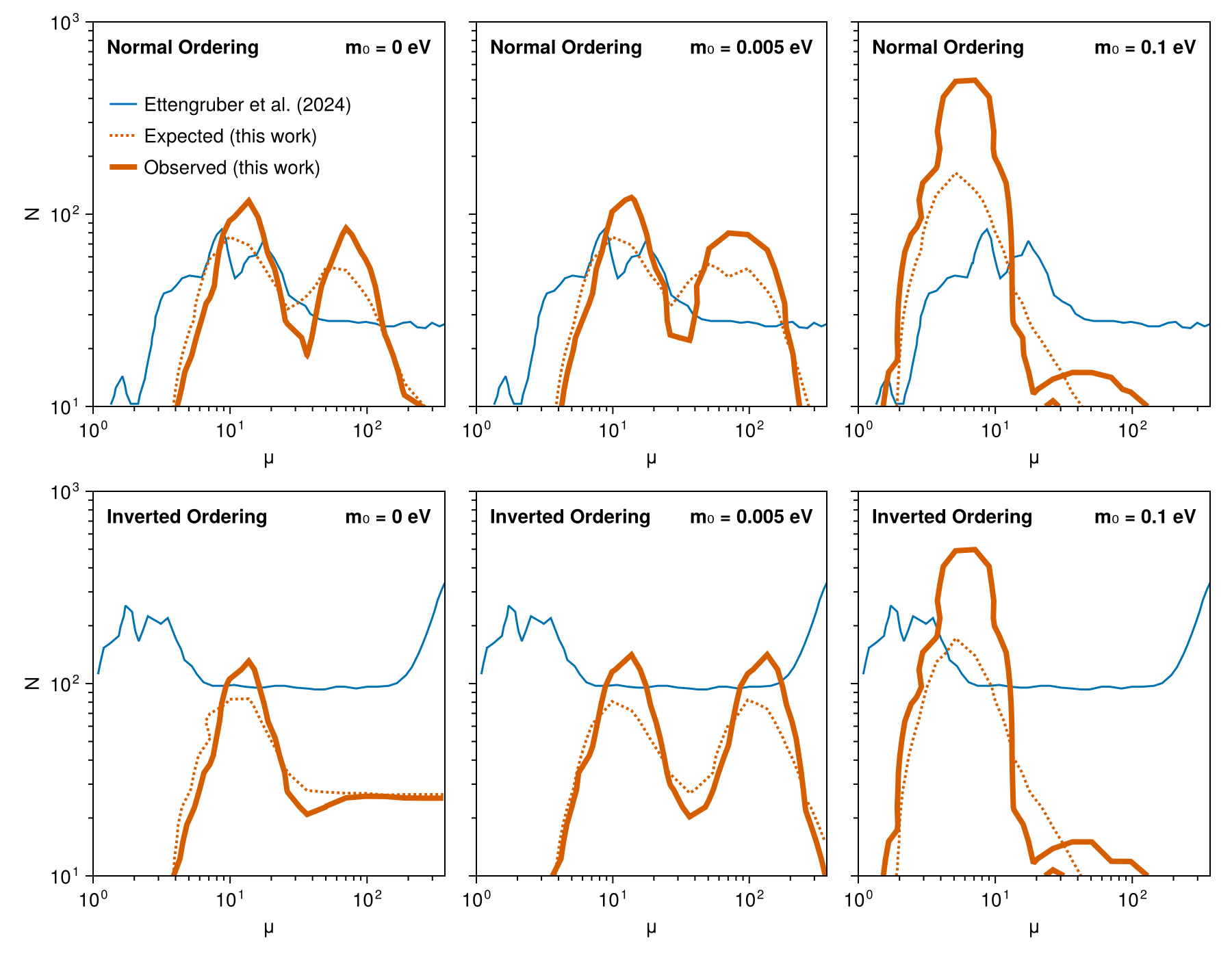}
    \caption{Lower limits on the number of SM Copies. The regions underneath the lines represent values excluded at the 99\% confidence levels, shown for normal (top row) and inverted (bottom row) mass ordering and different choices of the lightest neutrino mass $m_0$: $m_0 = 0$~eV (left), $m_0 = 0.005$~eV (middle), $m_0 = 0.1$~eV (right). The solid red lines show our observed limits, while the dotted red lines show our median expected sensitivities. The blue lines compare with existing neutrino bounds~\cite{ettengruber2024testing} (which are $m_0$-independent).}
    \label{fig:smcopies_limits}
\end{figure*}
The corresponding constraints on the SM copies framework are shown in Fig.~\ref{fig:smcopies_limits}, where exclusion regions are presented in the $(N, \mu)$ parameter space for representative values of the lightest neutrino mass $m_0$. We obtain lower limits of up to
\begin{equation*}
N \gtrsim \mathcal{O}(400),
\end{equation*}
depending on the assumed value of $m_0$ and the mass ordering.

The strongest constraints are achieved in regions of $\mu$ that bring the heavy states to the eV scale, for which the matter-enhanced resonances fall within the energy range of the sample. 
Some of the limits feature an interesting double-peaked structure, which is due to distinct resonances (see also Fig.\ref{fig: comparison flux ratio models}) that move in and out of the sensitive region when scaling $\mu$. In principle, we expect three resonances in the spectrum, but depending on the choice of light neutrino mass $m_0$, the mass ordering, and $\mu$, some resonances vanish or become degenerate. For example, for $m_0=0$\,eV, the lowest mass resonance vanishes, and in the case of inverted ordering the two higher mass resonances merge into one. This is the reason for the single vs. double peak structure in the two leftmost panels in Fig.\ref{fig:smcopies_limits}. For the rightmost panels, i.e., large $m_0$, the mass splittings become comparatively small, and all resonances merge into a degenerate one, leaving only a single peak that is quasi-independent of the mass ordering. The small kink visible for $m_0=0.1$\,eV at a mass factor $\mu\approx20$ is due to a closed 90\% C.L. at $N\approx20$.
This closed contour arises because the data in that energy-zenith region mildly prefer a small deficit; its significance is well below
$3\sigma$, and it does not constitute evidence for a signal.
Translating this region into the 3+1 model parameter space results in a sterile mass of around 2\,eV and a mixing of $\sin^2 2\theta \approx 0.2$, the same region reported in \cite{IceCubeCollaboration:2024nle}.

In parts of the parameter space, our limits constitute the most stringent terrestrial constraints to date, while in other regions they complement existing bounds from oscillation experiments. This is also illustrated in Fig.~\ref{fig:smcopies_limits}, where we compare the limits obtained in this work to the constraint of~\cite{ettengruber2024testing}.

\section{CONCLUSIONS \label{sec: conclusions}}

We have searched for signatures of low-scale gravity scenarios in 10.7 years of high-energy atmospheric neutrino data collected with the IceCube Neutrino Observatory. In both the ADD framework and the scenario with multiple Standard Model copies, the presence of additional sterile states would induce characteristic, energy- and zenith-dependent distortions through matter-enhanced active--sterile transitions. No such deviations from the three-flavor expectation are observed in the data. 

The absence of these signatures allows us to place strong constraints on the compactification radius in the ADD scenario, significantly restricting the allowed parameter space of large extra dimensions. In the framework with multiple Standard Model copies, we derive constraints that probe previously unexplored regions of parameter space. These results demonstrate that high-energy neutrino propagation through the Earth provides direct sensitivity to low-scale gravity realizations in the neutrino sector. Future improvements in statistics and reconstruction are expected to further strengthen these constraints and extend the reach of such searches.

\begin{acknowledgements}
The authors gratefully acknowledge the support from the following agencies and institutions:
USA {\textendash} U.S. National Science Foundation-Office of Polar Programs,
U.S. National Science Foundation-Physics Division,
U.S. National Science Foundation-EPSCoR,
U.S. National Science Foundation-Office of Advanced Cyberinfrastructure,
Wisconsin Alumni Research Foundation,
Center for High Throughput Computing (CHTC) at the University of Wisconsin{\textendash}Madison,
Open Science Grid (OSG),
Partnership to Advance Throughput Computing (PATh),
Advanced Cyberinfrastructure Coordination Ecosystem: Services {\&} Support (ACCESS),
Frontera and Ranch computing project at the Texas Advanced Computing Center,
U.S. Department of Energy-National Energy Research Scientific Computing Center,
Particle astrophysics research computing center at the University of Maryland,
Michigan State University,
Astroparticle physics computational facility at Marquette University,
NVIDIA Corporation,
and Google Cloud Platform;
Belgium {\textendash} Funds for Scientific Research (FRS-FNRS and FWO),
FWO Odysseus and Big Science programmes,
and Belgian Federal Science Policy Office (Belspo);
Germany {\textendash} Bundesministerium f{\"u}r Forschung, Technologie und Raumfahrt (BMFTR),
Deutsche Forschungsgemeinschaft (DFG),
Helmholtz Alliance for Astroparticle Physics (HAP),
Initiative and Networking Fund of the Helmholtz Association,
Deutsches Elektronen Synchrotron (DESY),
and High Performance Computing cluster of the RWTH Aachen;
Sweden {\textendash} Swedish Research Council,
Swedish Polar Research Secretariat,
National Academic Infrastructure for Supercomputing in Sweden (NAISS),
and Knut and Alice Wallenberg Foundation;
European Union {\textendash} EGI Advanced Computing for research;
Australia {\textendash} Australian Research Council;
Canada {\textendash} Natural Sciences and Engineering Research Council of Canada,
Calcul Qu{\'e}bec, Compute Ontario, Canada Foundation for Innovation, WestGrid, and Digital Research Alliance of Canada;
Denmark {\textendash} Villum Fonden, Carlsberg Foundation, and European Commission;
New Zealand {\textendash} Marsden Fund;
Japan {\textendash} Japan Society for Promotion of Science (JSPS), Ministry of Education, Culture, Sports, Science and Technology (MEXT), and Institute for Global Prominent Research (IGPR) of Chiba University;
Korea {\textendash} National Research Foundation of Korea (NRF);
Switzerland {\textendash} Swiss National Science Foundation (SNSF).
\end{acknowledgements}

\bibliographystyle{unsrt}
\bibliography{mybib}

\end{document}